\documentclass[12pt]{iopart}

\usepackage{graphicx}
\usepackage[latin1]{inputenc}
\usepackage{hyperref}
\begin{document}

\title{Wave packet dynamics of K$_2$ attached to helium nanodroplets}

\author{P. Claas$^1$, G. Droppelmann$^1$, C. P. Schulz$^2$, M. Mudrich$^3$, F. Stienkemeier$^3$}

\address{$^1$Fakult\"at f\"ur Physik, Universit\"at Bielefeld, 33615 Bielefeld, Germany}
\address{$^2$Max-Born-Institut, Max-Born-Strasse 2a, 12489 Berlin, Germany}
\address{$^3$Physikalisches Institut, Universit\"at Freiburg, 79104 Freiburg, Germany}
\ead{mudrich@physik.uni-freiburg.de}

\begin{abstract}
The dynamics of vibrational wave packets excited in K$_2$ dimers attached to superfluid
helium nanodroplets is investigated by means of femtosecond pump-probe spectroscopy. The
employed resonant three-photon-ionization scheme is studied in a wide wavelength range
and different pathways leading to K$^+_2$-formation are identified. While the wave packet
dynamics of the electronic ground state is not influenced by the helium
environment, perturbations of the electronically excited states are observed. 
The latter reveal a strong time dependence on the timescale $3-8$\,ps which directly
reflects the dynamics of desorption of K$_2$ off the helium droplets.
\end{abstract}


\maketitle

\section{Introduction}

The advent of femtosecond (fs) spectroscopy has allowed the visualization of the
nuclear motion in molecular systems in
``real-time'', delivering new insight into the
nature of the chemical bond and of the dynamics of chemical
reactions~\cite{Zewail:2000,Zewail:1994,Manz:1995,Chergui:1995}. In femtosecond
pump-probe experiments this visualization relies on the creation and
interrogation of vibrational wave packets, i.~e.~coherent superpositions of vibrational
states excited by the fs laser pulse. Fs pump-probe spectroscopy has become a
widely-spread method providing complementary information to spectroscopy in the frequency
domain because fast dynamical processes are often directly accessible.

Besides fs studies of molecules prepared in the gas phase, the real-time spectroscopy of
molecules isolated in inert noble-gas matrices is an active field of research. On the one
hand, matrix isolated molecules display a large variety of manybody phenomena such as
solvational shifts of electronic states, electron phonon coupling, matrix induced
electronic and vibrational relaxation and decoherence, reduced dissociation probability
(``caging effect''), excimer and exciplex formation, charge
recombination~\cite{Apkarian:1999}. On the other hand, these conceptually simple model
systems can still be handled theoretically.

The field was pioneered by Zewail and co-workers using I$_2$ as a molecular probe and
varying the structural properties of the solvent environment from gas~\cite{Wan:1997},
over clusters~\cite{Liu:1993} and liquid~\cite{Lienau:1994} to solid~\cite{Liu:1993}.
For example, in experiments with I$_2$ in Ar clusters Liu and coworkers found that the coherent
wave packet motion can survive the dissociation and matrix-induced recombination process~\cite{Liu:1993}.
Schwentner and cowokers have extended
real-time studies to halogen molecules (I$_2$, Cl$_2$, Br$_2$, ClF) in rare gas matrices, adding
several new aspects in this kind of studies \cite{Bargheer:1999,Fushitani:2006,Guehr:2004,Bargheer:2002}.
Besides, Hg$_2$ has been investigated in a cryogenic Ar
matrix~\cite{Gonzalez:2002}. Conservation of vibrational coherence during a few
picoseconds despite matrix induced electronic relaxation was observed. Due to the
difficulty of implanting impurities into liquid helium, no real-time measurements with
molecules in bulk liquid or solid helium have been reported until now. However,
femtosecond laser-induced ionization of liquid helium and the subsequent preparation of
He$_2^*$ excimers in superfluid He has been reported~\cite{Benderskii:1999}. Recently,
electronic coherence relaxation of I$_2$ in helium and other gases has been measured by
fs photon echo~\cite{Comstock:2003}. 

Among the first molecules to be studied by fs spectroscopy in the gas-phase were alkali
dimers~\cite{Papanikolas:1995,Baumert:1991,
Baumert:1992,Riedle:1995,Riedle:1996,Rodriguez:1993,Blanchet:1995}. Cold alkali dimers
are easily prepared in a molecular beam and strong electronic transitions are well
accessible by common fs laser sources. The potassium dimer K$_2$ has been studied in
detail in a series of one and two-color experiments demonstrating the propagation of
vibrational wave packets in the ground state and in different excited
states~\cite{Riedle:1995, Riedle:1996, Rutz:1996, Rutz:1997, Schwoerer:1997, Pausch:1999,
Nicole:1999}. By varying the excitation laser wavelength and intensity, different
pathways leading to ionization were identified. In addition, more subtle effects were
studied such as the effect of spin-orbit coupling on the nuclear dynamics in the
isotopomer $^{39,39}$K$_2$ which is absent in $^{39,41}$K$_2$~\cite{Rutz:1996}. Visualizing
the time evolution of vibrational frequencies using spectrograms was found to be useful
for identifying transient phenomena such as fractional revivals in one-color pump-probe
scans which are absent in the case of two-color two-photon ionization~\cite{Rutz:1997}.

In this work we present the first fs experiments probing wave packet propagation of molecules
attached to superfluid helium nanodroplets. Today, helium nanodroplets are widely applied
as a nearly ideal cryogenic matrix for spectroscopy of embedded molecules and as
nanoscopic reactors for building specific molecular
complexes~\cite{Toennies:2004,Stienkemeier:2006}. Alkali atoms and molecules represent a
particular class of dopant particles due to their extremely weak binding to helium
droplets with binding energies on the order of 10\,K (7\,cm$^{-1}$). From both theory and experiment it
is known that alkali dimers reside in bubble-like structures on the surface of helium
droplets~\cite{Dalfovo:1994,Ancilotto:1995,Stienkemeier2:1995}. Therefore, spectroscopic
shifts of the electronic excitation spectra of alkali dimers are in the range of only a few cm$^{-1}$ with respect to the gas
phase~\cite{Stienkemeier:1995,Stienkemeier2:1995,Higgins:1998,HigginsThesis:1998,Higgins:2000}.
Upon electronic excitation, alkali atoms and molecules mostly desorb from the droplets as
a consequence of evaporation of helium atoms after energy deposition in the helium
droplet ~\cite{Stienkemeier:1996,Callegari:1998,Mudrich:2004}. The time scale for the
desorption process is largely unknown. Solely in an early time-resolved study the
desorption time of an alkali-helium excimer formed upon electronic excitation is
estimated to be much shorter than the fluorescence lifetime~\cite{Reho2:2001}.

Recently, the dynamics of photodissociation of molecules inside helium droplets has been
studied both theoretically and experimentally. In Ref.~\cite{Takayanagi:2003},
photodissociation of Cl$_2$ embedded in helium clusters is studied by numerical
simulation revealing reduced caging effects due to quantum properties of helium. The
translational dynamics of CF$_3$ in liquid helium was investigated experimentally by
photodissociation of CF$_3$I dissolved in helium droplets~\cite{Braun:2004}. Using ion
imaging techniques, the velocity distribution of the fragments escaping from the droplets
was found to be considerably shifted to lower speeds. This behavior was successfully
modeled assuming binary collisions with helium atoms. The photoionization dynamics of
aniline doped helium droplets was investigated by photoelectron
spectroscopy~\cite{Loginov:2005}. In these experiments, droplet-size dependent lowering
of the ionization threshold upon solvation was found.

Due to their weak coupling to the surrounding helium environment alkali dimers on helium
droplets can be viewed as an intermediate system between free molecules and molecules
isolated in conventional cryogenic matrices. The wave packet propagation is expected to
be only weakly perturbed and new dynamical phenomena resulting from the desorption of the
dimers upon fs excitation are observable. The highly quantum nature of superfluid $^4$He
droplets vs. normal fluid $^3$He droplets may affect the coupling of the wave packet
motion to the droplet~\cite{Grebenev:1998,HartmannPRL:1996}. Moreover, exotic high-spin
molecules and complexes formed on He droplets can be studied in real
time~\cite{Higgins:1998,Higgins:1996,Schulz:2004}. The wave
packet dynamics of Na$_2$ in the lowest triplet state $b^3\Sigma_u^+$ will be presented
in a separate paper.

\section{Experimental}

\begin{figure}[t]
\begin{center}
\includegraphics* [scale=0.6]{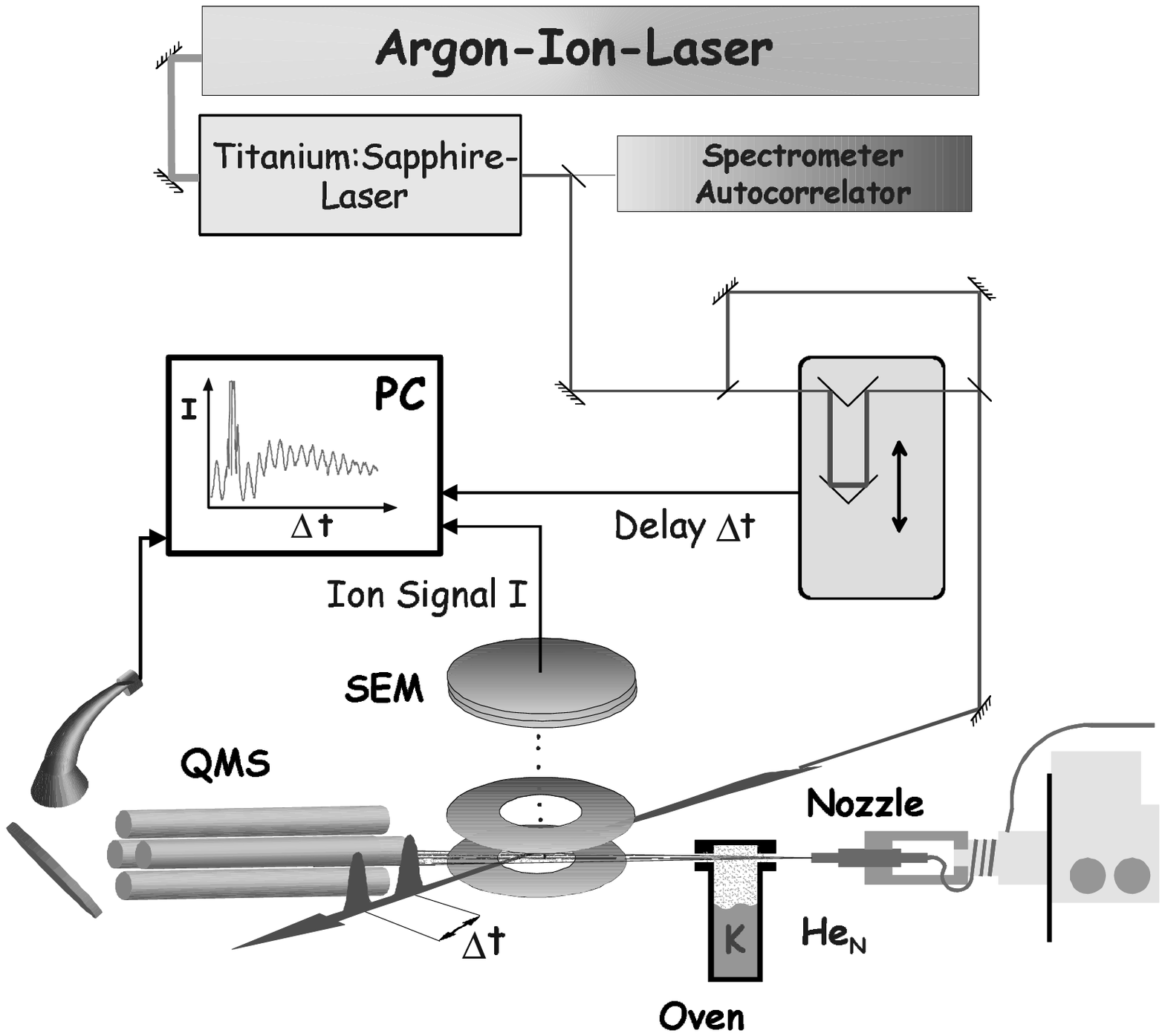}
\caption{Schematic illustration of the experimental setup.
\label{fig:setup} }
\end{center}
\end{figure}

The experimental setup consists of three main parts, as illustrated in
Fig.~\ref{fig:setup}: A molecular beam line for ${K_{2}}$-doped helium droplets, an
optical system to produce time delayed femtosecond laser pulses and a mass selective
detection unit. The beam line consists of three differentially pumped vacuum chambers. In
the first chamber the helium droplet beam is produced by condensation of helium gas
(purity 99.9999\%) in a supersonic expansion from a cold nozzle with a diameter
$d=10\,\mu$m into vacuum. At a nozzle temperature $T\approx 19$\,K and a stagnation
pressure $p\approx 50$\,bar the droplets contain about 5000 atoms. By varying the
temperature of the nozzle the average size of the droplets can be changed.
To maintain a background pressure of $\approx 10^{-3}$\,mbar we use a 8000\,l/s oil
diffusion pump backed up by a roots and rotary vane pump.

The beam enters the second chamber through a skimmer with an aperture of 400$\,\mu$m
diameter. There the droplets pass a heated cell where they successively pick up two
potassium atoms. Due to their high mobility on the superfluid helium droplets the
picked-up atoms bind together to form dimers which are weakly bound in a dimple on the
surface of the the helium nanodroplets. Collisional as well as binding energy is
dissipated by the helium droplet through evaporation of helium atoms, which may cause the
desorption of the alkali dimers from the droplets. Since the amount of internal energy
released upon formation of ground state ($X^1\Sigma_g^+$) dimers greatly exceeds the one
released upon formation of dimers in the lowest triplet state $a^3\Sigma_u^+$, the later
has a higher chance to remain attached to the droplet. This leads to an enrichment of the
droplet beam with high-spin dimers and clusters compared to covalently bound systems
\cite{Higgins:1998,Schulz:2004}. Nevertheless singlet ground state alkali dimers on
helium droplets have been studied~\cite{Stienkemeier2:1995,Higgins:2000}. The
alkali-helium droplet complex eventually equilibrates at the terminal temperature of pure
helium droplets of 380\,mK. Thus, only the lowest vibrational state $v''=0$ and a few
rotational states are populated which provides well-defined starting conditions for a
pump-probe experiment. The number of collisions of the droplets with free dopant atoms
inside the vapor cell is controlled by the oven temperature. Given the flight distance of
the droplets of 1\,cm inside the vapor cell the potassium reservoir is
heated to a temperature $T\approx 410\,$K to achieve highest probability for pick-up of
two dopant atoms per droplet.

In the third vacuum chamber the doped droplet beam is intersected by the
laser beam at right angle. The femtosecond laser pulses are generated by
an Ar$^{+}$-laser pumped Ti:Sapphire laser (Tsunami, Spectra-Physics)
at 80\,MHz repetition rate and at an average
power output between 1.2 and 1.4\,W depending on the wavelength.
This corresponds to a pulse energy of about 16\,nJ. The pulses have a
duration of $\approx 110$\,fs and a spectral
bandwidth of $\approx 135$\,cm$^{-1}$ (FWHM).

\begin{figure}[t]
\begin{center}
\includegraphics* [scale=0.6]{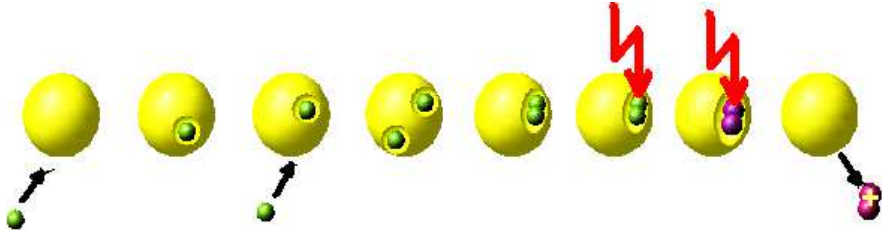}
\caption{(Color online) Schematic illustration of the time sequence of the experiment. Successive pick-up
of two alkali atoms by the helium droplet is followed by formation of a dimer. The later
is excited by the pump pulse and ionized by the probe pulse.
\label{fig:dimer}
}
\end{center}
\end{figure}

The laser beam is sent through a Mach-Zehnder interferometer to split the pulses into two
time delayed pump and probe parts with equal intensity. The time delay is controlled by a
commercial translation stage (PI M531DG) having 0.225\,fs as the smallest time increment.
The collinear pulse pairs are then focused into the doped helium droplet
beam using a lens of 150\,mm focal length. The interaction volume is located inside the
formation volume of a quadrupole mass spectrometer to mass selectively detect the
photoions. We estimate the intensity of the laser light at the crossing point to be
$\approx 0.5\,$GW/cm$^{2}$ which corresponds to moderate intensity also used in the
experiments reported in Refs.~\cite{Riedle:1995,Nicole:1999}. The time sequence of the
experiment is illustrated schematically in Fig.~\ref{fig:dimer}.

\begin{figure}[t]
\begin{center}
\includegraphics* [scale=1]{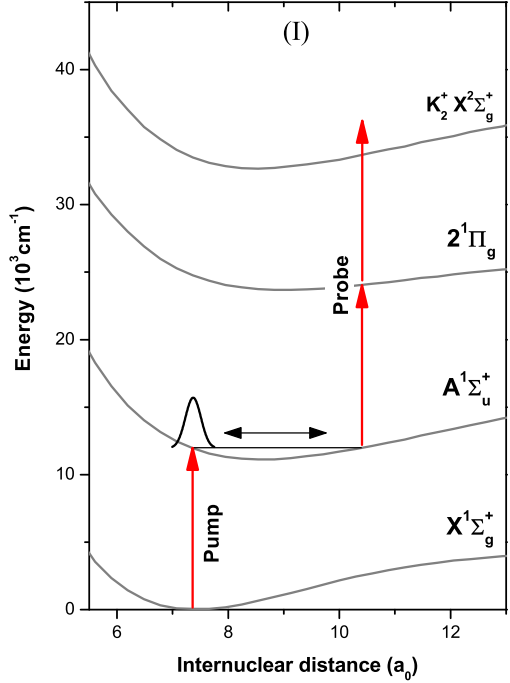}
\caption{(Color online) Potential energy surfaces of neutral K$_2$ and of K$_2^+$
populated in the femtosecond pump-probe experiment. The arrows indicate pump and probe transitions for $\lambda = 833\,$nm.
\label{fig:scheme1}
}
\end{center}
\end{figure}

\section{Pump-probe-spectroscopy of K$_2$}

\begin{figure}[t]
\begin{center}
\includegraphics* [scale=1]{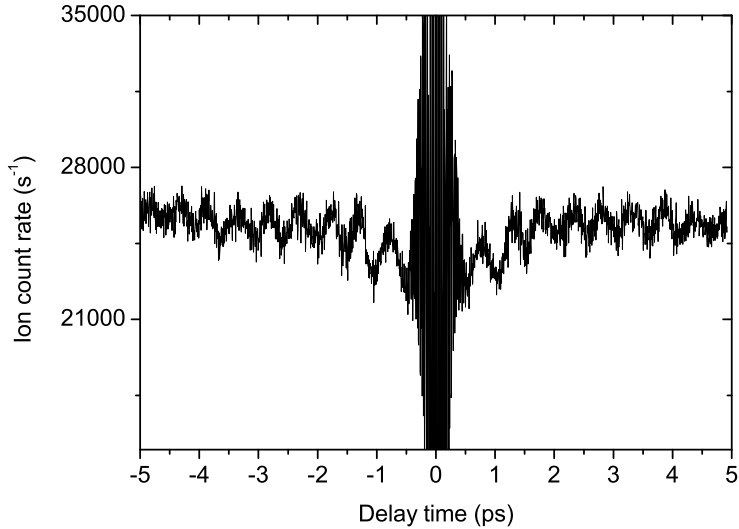}
\caption{Raw K$_2^+$ ion signal after pump-probe photoionization of K$_2$ on helium nanodroplets as a function of delay time between pump and probe pulses recorded at $\lambda =810\,$nm.
\label{fig:raw}
}
\end{center}
\end{figure}

\begin{figure}[t]
\begin{center}
\includegraphics* [scale=1]{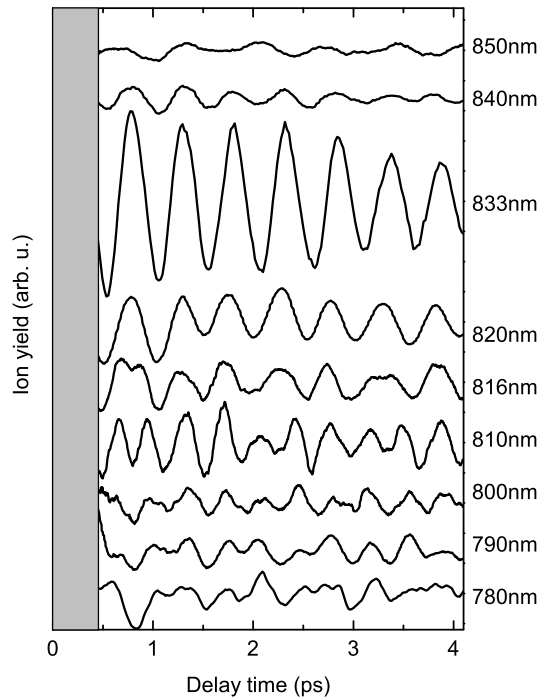}
\caption{Processed pump-probe scans of K$_2$ on helium nanodroplets for various 
excitation laser wavelengths $\lambda$.
\label{fig:ppspectra}
}
\end{center}
\end{figure}

The one-color pump-probe experiments reported in this work are carried
out in the wavelength region $\lambda =780-850$\,nm.
The excitation scheme leading to wave packet motion and subsequent ionization is
illustrated in Fig.~\ref{fig:scheme1} using the gas-phase potentials from
Refs.~\cite{Magnier:1996,Lyyra:1992}. The relevant potential energy curves are the ground
state $X^1\Sigma_g^+$, the first excited state $A^1\Sigma_u^+$, the doubly excited state
$2^1\Pi_g$ and the ground state of the ion K$_2^+$, $X^2\Sigma_g^+$. The arrows indicate
the excitation pathway leading to photoionization of K$_2$ at $\lambda =833\,$nm. Starting from the
vibrational ground state $X^1\Sigma_g^+$, $v''=0$, absorption of one photon from the pump
pulse creates a coherent superposition of ca. 5 vibrational states around $v'=12$ in
the $A$-state at the classical inner turning point. The vibrational wave packet then
propagates outwards and is transferred to the ionic state in a two-photon step via the
intermediate resonant state $2\Pi$ according to the resonance condition that photon
energy equals the difference in potential energies of $A$ and $2\Pi$ states.
It is essential to note that the $2\Pi$-state
opens a Franck-Condon (FC) window for resonance enhanced ionization
around a well-defined internuclear distance and therefore filters
out the wave packet dynamics in the $A$-state~\cite{Riedle:1995}. Thus, the measured photoionization yield shows an oscillatory
structure corresponding to the vibrational motion of the molecules
(cf.~Figs.~\ref{fig:raw}, \ref{fig:ppspectra}). This oscillation has a well-defined phase
$\Phi$ at zero delay times. At 833\,nm the resonance condition is fulfilled at the outer
turning point of the $A$-state. Thus, the time between wave packet creation and
ionization roughly equals half a period of the wave packet oscillation and the detected
signal has a phase $\Phi^{A}_{833}\approx\pi$. 

The described scheme at 833\,nm is
particularly favorable for several reasons. First, significant population of the
2$^1\Pi$-state during the pump-pulse is suppressed because the $2\Pi\leftarrow A$
transition is stronger than the $A\leftarrow X$ transition~\cite{Riedle:1995}. This leads
to pure (1+2)-photon excitation without contributions of (2+1)-photon transitions.
Second, the outer turning points on the potential curves of the $A$ and the $X$-state
coincide. Therefore the FC factors for the $A\leftarrow X$ transition is
optimal at this wavelength~\cite{Rutz:1997}. Moreover, since the dynamics in the
$A$-state is sampled at the turning point the detected oscillation directly reflects the
nuclear dynamics~\cite{Nicole:1999}.

\begin{figure}[t]
\begin{center}
\includegraphics* [scale=1]{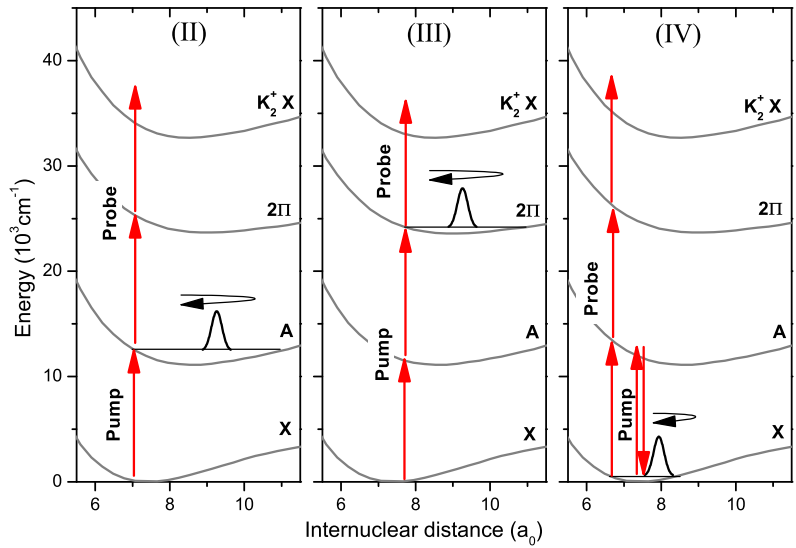}
\caption{(Color online) Schematic illustration of different pump-probe mechanisms leading to photoionization of K$_2$ (see text).
\label{fig:scheme2}
}
\end{center}
\end{figure}

Besides the described ionization scheme (type I) in the K$_2$ experiments, the following
alternative pathways are depicted in Fig.~\ref{fig:scheme2}:
\begin{itemize}
\item{Type II: Wave packet excitation in the $A$-state. The transition to the
ionic state takes place at the inner turning point, which results in a phase
$\Phi^{A}_{780}\approx 2\pi$. This scheme applies to the excitation at $\lambda\sim 780\,$nm.}

\item{Type III: Wave packet excitation in the $2\Pi$-state by absorption of two
photons from the pump pulse. Ionization occurs by one-photon step during the probe
pulse. This schemes is active at $\lambda\sim 780-800\,$nm.}

\item{Type IV: Wave packet excitation in the $X$-ground state by resonant impulsive stimulated
Raman scattering (RISRS) during the pump pulse~\cite{Riedle:1996}. Transition to the
ionic state is achieved by absorption of three photons from the probe pulse. This scheme
applies to the excitation at $\lambda\approx 780-820\,$nm.}
\end{itemize}

Around 850\,nm, the excitation proceeds similarly to the type III scheme but involving the
doubly excited state $4^1\Sigma_g^+$ instead of the $2\Pi$-state.
Which one of the mentioned schemes is active, depends on the laser wavelength, intensity
and polarization. For example in the one-color pump-probe experiments with K$_2$ in a
molecular beam  reported by Nicole \textit{et al.}~\cite{Nicole:1999}, it was shown that
type~I is the dominant process in the wavelength range $\lambda=837-816\,$nm, whereas at
$\lambda=800-779\,$nm type III also contributes. At $\lambda=779\,$nm the observed
transient was best reproduced by a sum of processes of type II and III. By increasing the
laser intensity up to 5\,GW/cm$^2$, it was shown that RISRS (type IV) significantly
contributes, which gives a parameter to control the specific ionization
pathway~\cite{Riedle:1996}. A different method for generating a wave packet in the ground
state was studied using a two-color pump-dump-probe scheme involving the
$B$-state~\cite{Pausch:1999}. In that work, the population of the ground state wave
packet was controlled by varying the pump-dump delay and to a limited extent by varying
the pump and dump laser wavelengths. Recently, Brixner \textit{et al.} have shown that
laser polarization opens an additional level of quantum control over the ionization
mechanism. By altering the mutual polarization of pump and probe pulses using an optimal
control scheme they were able to selectively enhance type I or type III transitions.

A typical real-time pump-probe spectrum of K$_2$ attached to helium nanodroplets recorded
at the mass of the $^{39,39}$K$_2^+$ dimer at $\lambda = 820\,$nm is exemplified in
Fig.~\ref{fig:raw}. The signal is nearly symmetric with respect to the delay time zero
which results from the almost identical pump and probe pulses interchanging their roles at time zero. Within
the delay time window $\pm 0.5\,$ps the ion signal displays fast oscillations due to the
autocorrelation of pump and probe pulse. This time interval is masked out in the further
data analysis. For longer delay times, the ion signal features an oscillation on the ps
time scale reflecting the wave packet dynamics. This oscillation is biased by a large
background ion count rate which reduces the contrast to about 10\,\%. The background ion
rate results from 3-photon ionization (3PI) of K$_2$ during the pump or the probe pulse. A
second background contribution comes from K$_2$ dimers interacting with more than one
pump-probe pulse pair as they fly through the laser focus. Due to the short
period between two consecutive pump-probe pulse pairs of 12.5\,ns (1/80\,MHz) and the
comparable $\sim 10\,$ns lifetime of the excited states, the
illuminated  K$_2$ beam always contains a portion of already excited
dimers.

In addition to the constant background another prominent feature shown in
Fig.~\ref{fig:raw} is the increase
of the ion signal during the first 3\,ps. This is due to potassium clusters K$_{N>2}$
formed on the helium droplets being photoionized and partially fragmenting into K$^+_2$
fragment ions. Thus the fragmentation dynamics of alkali clusters is also accessible
in our experiments but will be discussed in a separate paper. In order to prepare the
data for Fourier transformation (FT) a band-pass filter is applied which suppresses the
constant offset and slow variations as well as high frequency noise.

The processed experimental transient ion signals are summarized in
Fig.~\ref{fig:ppspectra} for short delay times $0-4\,$ps. As expected from the gas-phase,
the most prominent oscillation occurs at $\lambda = 833\,$nm. Clearly, the oscillation
amplitude is damped within the shown time interval. As the wavelength decreases the
amplitudes quickly drop and a new higher frequency component appears which is most
apparent at $\lambda = 810\,$nm. For even smaller wavelengths, the transients feature a
more complicated structure, indicating that more than one frequency component
contributes. At $\lambda = 850\,$nm, the ionization trace is modulated by a slow
oscillation with a rather weak amplitude. This oscillation has the opposite phase as the
one observed at $\lambda = 833\,$nm.

\begin{figure}[t]
\begin{center}
\includegraphics* [scale=1]{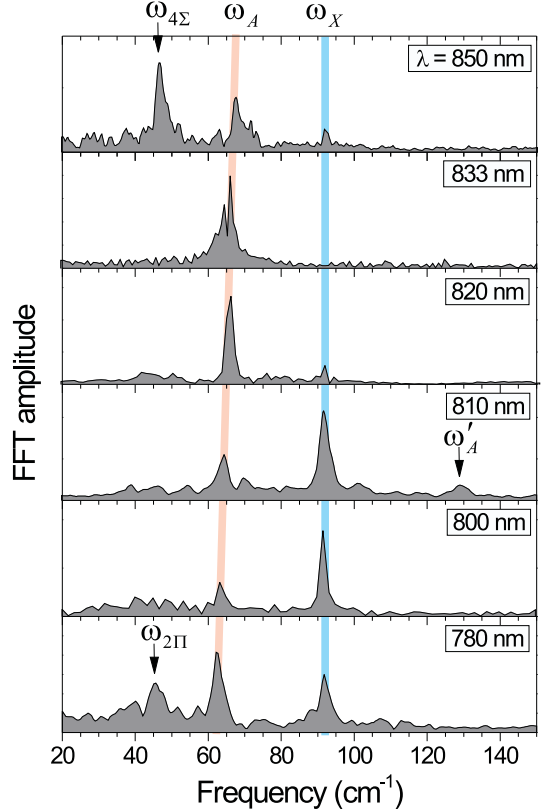}
\caption{(Color online) Fourier transforms of experimental pump-probe scans at different
laser wave lengths $\lambda$. The arrows and the colored straight lines indicate different
frequency components.
\label{fig:fft}
}
\end{center}
\end{figure}

In order to obtain more quantitative information about the exact frequencies and relative
amplitudes of the contributing frequency components the pump-probe traces are Fourier
transformed in the entire recorded delay time interval and plotted in Fig.~\ref{fig:fft}.
For the sake of clarity, the spectra are individually normalized to their maxima. One clearly distinguishes
two characteristic frequencies which appear in all spectra. The frequency $\omega_X
=92.1(7)\,$cm$^{-1}$ matches the gas-phase vibrational constant
$\omega_e=92.4\,$cm$^{-1}$ of the ground state $X^1\Sigma_g^-$~\cite{Magnier:1996}. Thus,
a wave packet composed of the vibrational states $v''=0,\, 1$ is excited by RISRS (type IV
transition). This type of ionization process was observed in
Refs.~\cite{Riedle:1995,Riedle:1996} in one-color experiments. In a two-color
pump-dump-probe experiment wave packets around $v''=1,\,2$ were excited leading to a level
spacing $\omega_X(1,2) =91.5\,$cm$^{-1}$ \cite{Pausch:1999}. As expected for RISRS,
$\omega_X$ does not display any change with varying laser wavelength. Moreover, no
significant frequency shift due to the interaction with the helium droplet is observed.
This result is not surprising when comparing to vibrational spectroscopy of molecules
inside the droplets. Vibrational shifts typically lie in the range $0.1-2\,$cm$^{-1}$
\cite{Toennies:2004}. Since alkali dimers reside on the surface, even smaller shifts are
expected.

\begin{figure}[t]
\begin{center}
\includegraphics* [scale=3]{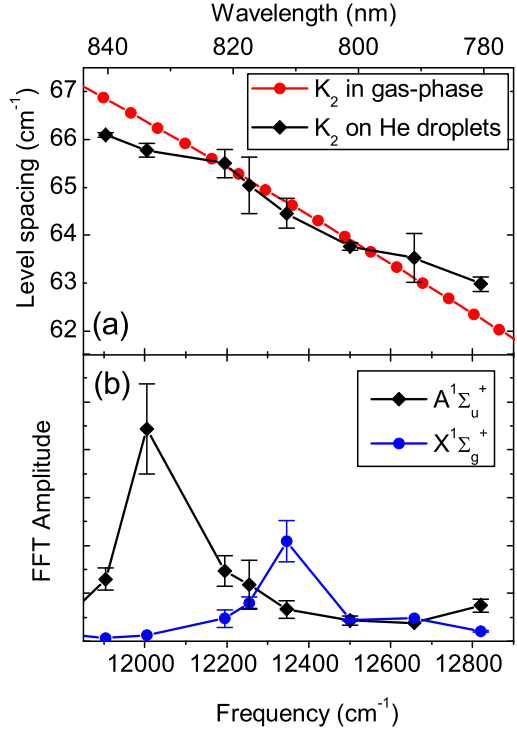}
\caption{(Color online) (a) Oscillation frequencies of wave packets in the $A$-state of K$_2$ attached
to helium droplets vs. gas-phase values from Ref.~\cite{Lyyra:1992}. (b) Relative
amplitudes of the FT components $\omega_A$ and $\omega_X$ as a function of laser
frequency. \label{fig:spacing} }
\end{center}
\end{figure}

The frequency component in the range 64-66\,cm$^{-1}$ belongs to the
vibration in the $A$-state. It clearly shifts to higher values with
increasing laser wavelength. This results from the anharmonicity of the $A$-state
potential: As the excitation wavelength decreases higher vibrational levels, having a
narrower energy spacing, are populated. The frequencies $\omega_A$ obtained from FT of
the first 7\,ps are plotted in Fig.~\ref{fig:spacing} (a). These frequencies are average
values of level spacings of about 5 vibrational levels which lie inside the spectral
profile of the fs laser. In order to compare this result to unperturbed K$_2$ dimers, the
corresponding average level spacings are calculated from experimental data from
Ref.~\cite{Lyyra:1992} which are plotted as full circles in Fig.~\ref{fig:spacing} (a).
For example the value close to $\lambda =810\,$nm (12350\,cm$^{-1}$) is calculated as the average of two level spacings,
$(G(v'+1)-G(v'-1))/2$, for $v'=18$. Here, $G(v')$ stands for the vibrational energy of level
$v'$. Without quantitatively analyzing the populations and the helium-induced shifts of
the $v'$-levels, Fig.~\ref{fig:spacing} (a) indicates a clear trend: In the wavelength
range $\lambda =840-780\,$nm the vibrational level spacings of K$_2$ on helium droplets
vary more slowly than the ones of free K$_2$. Thus, the helium droplet environment
influences the shape of the potential curve of the $A$-state of K$_2$ such that
anharmonicity is reduced. Note that the frequencies obtained from FT of our data in the
time interval $10-20\,$ps, i.\,e. when desorption is complete,
agree with the calculated values within the error bars.

The amplitudes of the frequency components of $X$ and $A$-states obtained from FT of the
first 7\,ps delay time are shown in Fig.~\ref{fig:spacing} (b). The maxima are located around
12350\,cm$^{-1}$ (810\,nm) and 12000\,cm$^{-1}$ (833\,nm), respectively. Surprisingly, in the range 800-815\,nm, the
$X$-component dominates over the $A$-component, which strongly differs from the gas-phase
observations at moderate laser intensities~\cite{Riedle:1995,Nicole:1999}. According
to~\cite{Riedle:1995,Nicole:1999}, wave packets in the $X$-state are not detected at moderate laser
intensities due to the mismatch of the FC window for ionization determined by
the $^2\Pi$-state and the $X$-state wave packet. In the case of K$_2$ on helium droplets, a well
localized FC window appears in 3PI from the $X$-state.
Because of the Pauli repulsion of the helium environment towards the alkali valence
electrons and the induced change in the spacial valence electronic distribution upon
electronic excitation, in particular the higher excited states are expected to evolve
larger energy shifts. The results demonstrate that the helium-induced shifts of the potential energy
curves significantly alter FC overlaps. In the following, several
characteristic wavelengths are discussed in more detail.

\paragraph{\textbf{833\,nm}}

\begin{figure}[t]
\begin{center}
\includegraphics* [scale=1]{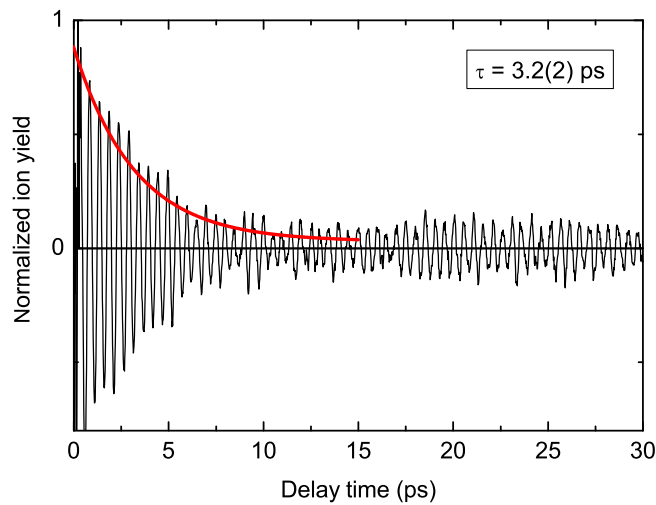}
\caption{(Color online) Transient ionization signal of K$_2$ attached to helium nanodroplets at
$833\,$nm. 
The thick lines shows the result of fitting an
exponential decay function with constant offset to the absolute amplitude of the observed
oscillation. \label{fig:Decay} }
\end{center}
\end{figure}

The damped oscillation recorded at 833\,nm is plotted for longer delay times in Fig.~\ref{fig:Decay}.
In addition to the experimental data the result of fitting an exponential decay function with a constant offset to the absolute oscillation amplitude
is represented as a thick line. The exponential drop by about one order
of magnitude has a time constant $\tau =3.2(2)\,$ps and the constant offset amounts to merely 5\,\% of the initial amplitude.
This initial amplitude decrease is not observed in gas-phase measurements~\cite{Rutz:1996,Nicole:1999}.
Generally, this dynamics can originate from two different mechanisms connected to the influence of the helium matrix.
First, matrix-induced effects such as vibrational relaxation may lead to decoherence of the wave packet~\cite{Guehr:2002}.
In this case, the K$_2$ dimers are stuck to the helium surface and the coherent oscillation is
damped and eventually completely vanishes. However, this is not the case in our experiment.
Second, additional dynamics is introduced by the effect of desorption of K$_2$ from the helium
surface, leading to a transient perturbation of the wave packet motion. Finally, both
of the mentioned effects may contribute to the observed dynamics. As discussed in more detail
in the last section, we interpret the observed exponential drop as the effect of
desorption of K$_2$ from the helium nanodroplets. The distortion of the potential curves
resulting from the interaction with the helium leads to an enhancement of resonant 3PI which
vanishes as K$_2$ moves away from the helium surface.

The FT spectrum at 833\,nm shown in Fig.~\ref{fig:fft} clearly indicates that only one
frequency component contributes to the wave packet dynamics. The oscillation period and
phase during the initial exponential decay are found to be $T_{A_1}^{833nm}= 507\,$fs
($\omega_{A_1} =$65.8\,cm$^{-1}$) and the phase equals half the oscillation period
($\Phi_{A}^{833nm}=0.99\,\pi$, $\hat{=} 255$\,fs). This confirms that at this
wavelength K$_2$ is ionized following the pathway of type I as in the gas-phase. At delay
times exceeding the desorption time the oscillation period is reduced to
$T_{A_2}^{833nm}= 503\,$fs ($\omega_{A_2} =$66.3\,cm$^{-1}$). This value is in good
agreement with the gas phase measurements of Ref.~\cite{Lyyra:1992}. In gas-phase measurements
with $^{39,39}$K$_2$ recording much longer scans, Rutz~\textit{et al.} were able to
resolve by FT each frequency spacing between the vibrational levels contributing to the
wave packet. At 833.7\,nm it was found that the FT spectrum is dominated by two
frequency components $G(v')-G(v'-1)$ and $G(v'+1)-G(v')$ for $v'=13$ due to spin-orbit
coupling between the $A$-state and the $b^3\Pi_u$-state~\cite{Rutz:1996}. The center of
gravity of the lines lies at about 66\,cm$^{-1}$ which is again in good agreement with our
findings. This suggests that in our experiment a wave packet consisting of about 5
vibrational components around the $v'=13$ vibrational state is excited.

The remaining oscillation after the initial damping 
features a slow periodic modulation with a period of about 26\,ps. The gas phase
measurement with the isotopomer $^{39,39}$K$_2$ also shows a pronounced beat structure
with a period of about 10\,ps, induced by the two dominating components mentioned above.
In the isotopomer $^{39,41}$K$_2$ spin-orbit coupling is not active in this region of
the $A$-state potential~\cite{Rutz:1996}. In that case, the beat structure is less
pronounced and more complicated with longer periods $>50\,$ps, resembling more the one
that we measure with $^{39,39}$K$_2$ on helium droplets. The fact that the beating
pattern after desorption does not precisely match the observed beating in the gas-phase
measurements is not totally understood. Presumably, we account differences in the used
pulse widths and shapes as well as differences in the wave packet formed under perturbed
conditions for this discrepancy.

\begin{figure}[t]
\begin{center}
\includegraphics* [scale=1]{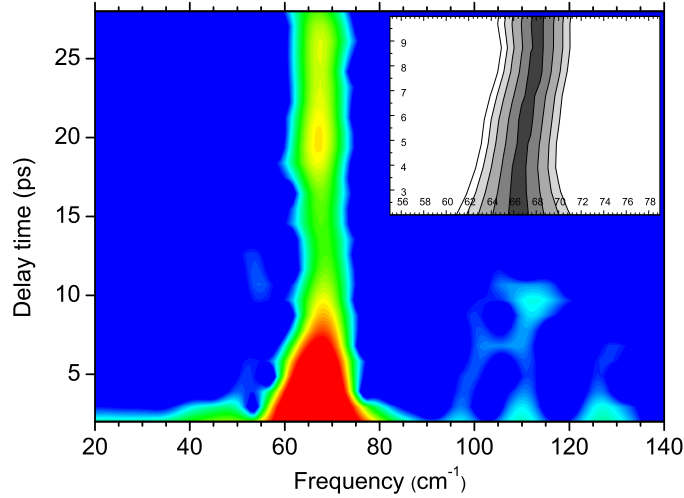}
\caption{(Color online) Logarithmic contour plot of the spectrogam calculated from the pump-probe
signals at the excitation wavelength $\lambda=833$\,nm. The inset extracts the first
10\,ps normalizing the data for each time step. \label{fig:spec833} }
\end{center}
\end{figure}

\begin{figure}[t]
\begin{center}
\includegraphics* [scale=1]{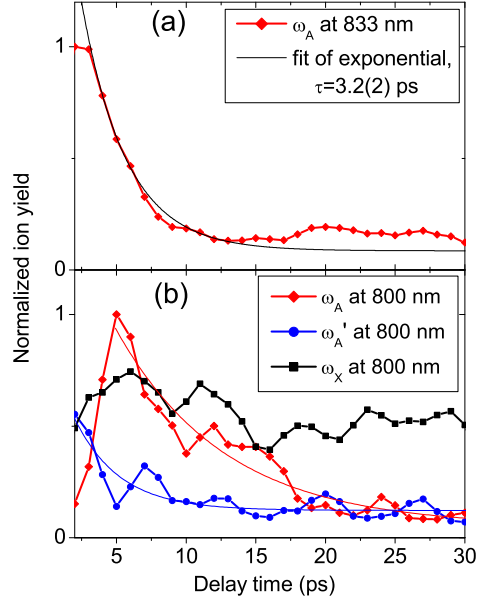}
\caption{(Color online) Time evolution of the relative amplitudes of different frequency components at
$\lambda =833\,$nm (a) and $\lambda =800\,$nm (b).
\label{fig:FTEvo}
}
\end{center}
\end{figure}

The evolution of $\omega_A$ can be followed in real-time by Fourier transforming the
data inside of a 3\,ps time window which slides across the entire data
set~\cite{Rutz:1997}. The resulting spectrogram is depicted in Fig.~\ref{fig:spec833} as
a contour plot with logarithmic scale. Naturally, the frequency resolution is reduced so
that $\omega^{833nm}_A(t_{delay})$ appears as a rather broad vertical band. The color scale goes
from blue for low to red for high amplitudes of the Fourier components. Thus, the initial
drop of the oscillation amplitude at $\omega^{833nm}_A\approx 66\,$cm$^{-1}$ is visualized as
intense red color which fades out at $t_{delay}\approx 10\,$ps and reappears at around
$t_{delay}\approx 20\,$ps indicating the revival mentioned above. The time evolution of the 
amplitude of $\omega^{833nm}_A$ obtained by averaging within a frequency interval 55-80\,cm$^{-1}$
is plotted in Fig.~\ref{fig:FTEvo} (a). Obviously, the data reproduce the decreasing
oscillation amplitude at 833\,nm as reflected by the fit curve from Fig.~\ref{fig:Decay} (solid line).

The shifting of the frequency $\omega^{833nm}_A$ is clearly visible in Fig.~\ref{fig:spec833}
as a deviation of the green band from a vertical line. In particular, during the
first 10\,ps in which the oscillation amplitude decreases exponentially, $\omega^{833nm}_A$
shifts roughly linearly to higher frequencies from 65.8\,cm$^{-1}$ to 66.3\,cm$^{-1}$, as
depicted in the inset of Fig.~\ref{fig:spec833} using a different color scaling. This
frequency shift is attributed to the effect of deminishing perturbation of K$_2$ by the helium
droplet as it desorbes off the droplet. Presumably, the perturbation of the $A$-state potential by helium
vanishes as K$_2$ desorbs from the helium surface thereby shifting the vibrational
frequencies back to their gas-phase values.

\paragraph{\textbf{820\,nm}}

The transient ionization signal at 820\,nm shown in Fig.~\ref{fig:ppspectra} resembles
the one at 833\,nm in the way that it features a pronounced oscillation which is damped
during the first 8\,ps. The period $T_A^{820nm}=520\,$fs ($\omega_A^{820nm}
=64.1\,$cm$^{-1}$) and phase $\Phi_A^{820nm}=0.96\,\pi$ (250\,fs) again indicate wave
packet propagation in the $A$-state (type I scheme). This result is somewhat surprising
since the corresponding gas-phase signals are quite different~\cite{Nicole:1999}. In the
gas phase, the maxima of the oscillation are split into two peaks of different
amplitudes, which is well described by the wave packet moving twice through the
FC window in each round trip. This is due to the fact that at 820\,nm the
outer turning point of the $2\Pi$-state (defining the FC window) no longer
coincides with the one in the $A$-state but is shifted to smaller internuclear distance.
This seems not to be the case when K$_2$ is attached to helium droplets. This can be
explained by either the shape of the $A$-state or of the $2\Pi$-state being influenced by
the helium such that the FC overlap localized at the outer turning points is
maintained down to a wavelength $\lambda = 816\,$nm. Since the vibrational frequencies in
the $A$-state are only weakly perturbed (Fig.~\ref{fig:spacing} (a)), most likely the
$2\Pi$-state is bent up to higher energies in the outer region of the potential well
similarly to the caging-effect of rare-gas matrices on the potential curves of I$_2$~\cite{Bargheer:1999}. 
Thus the the FC-window is shifted to shorter internuclear distances. 

\paragraph{\textbf{810\,nm}}

The new frequency component showing up at decreasing excitation wavelengths is most
apparent at $\lambda = 810\,$nm. We find a period $T_X^{810nm}=364\,$fs corresponding to
$\omega_X^{810nm} =91.7(8)\,$cm$^{-1}$ which is in good agreement with the gas-phase
vibrational constant $\omega_e=92.4\,$cm$^{-1}$ of the ground state
$X^1\Sigma_g^-$~\cite{Magnier:1996}. The phase is found to be
$\Phi_X^{810nm}=1.3\,\pi$ (245\,fs). Thus, we identify a type IV ionization sequence
in this wavelength region. This finding quite significantly differs from the gas-phase
results. According to~\cite{Riedle:1995,Nicole:1999} no wave packet dynamics in the $X$-state is
observed. Instead, wave packet excitation in the $2\Pi$-state contributes to an
increasing extent at shorter wavelengths. In the gas-phase, Fourier analysis of the
transients at 834\,nm does reveal a frequency component corresponding to wave packet
motion in the $X$-state. However the amplitude is negligible with respect to the
$A$-state contribution~\cite{Riedle:1995,Rutz:1996}. A dominating RISRS excitation was
only observed at laser intensities exceeding the ones
in our experiment by more than a factor 10~\cite{Riedle:1996}. This again demonstrates the sensitivity of
the population transfer and the produced wave packets on weak perturbations of the
involved states and the corresponding FC overlaps.

In addition to $\omega^{810nm}_X$ and $\omega^{810nm}_A$ there is a third component $\omega^{'810nm}_A\approx
128\,$cm$^{-1}$. This frequency is interpreted in terms of the wave packet in the
$A$-state being ionized via 2 distinct windows in the coordinate of internuclear
distance. The FC window is situated at the outer turning point at $\lambda =
833\,$nm and at the inner turning point at $\lambda = 780\,$nm. Apparently, the
transition between the two regimes takes place at $\lambda = 810\,$nm. The interpretation
of $\omega^{'810nm}_A$ being due to the coherent excitation of vibrational energy levels with
$\Delta v'=2$ as it was observed in Ref.~\cite{Rutz:1997} seems unlikely in our case. The
latter effect is a general feature depending only on the vibrational level spacing and
the spectral width of the laser pulses. Therefore, a doubled frequency $2\omega_A$ should
appear in all FT spectra, which is not the case in our measurements. One should note
here that the spectral width of the laser pulses in the experiments of
Ref.~\cite{Rutz:1997} were larger by a factor of 1.6 than in our experiments.

\begin{figure}[t]
\begin{center}
\includegraphics* [scale=2.5]{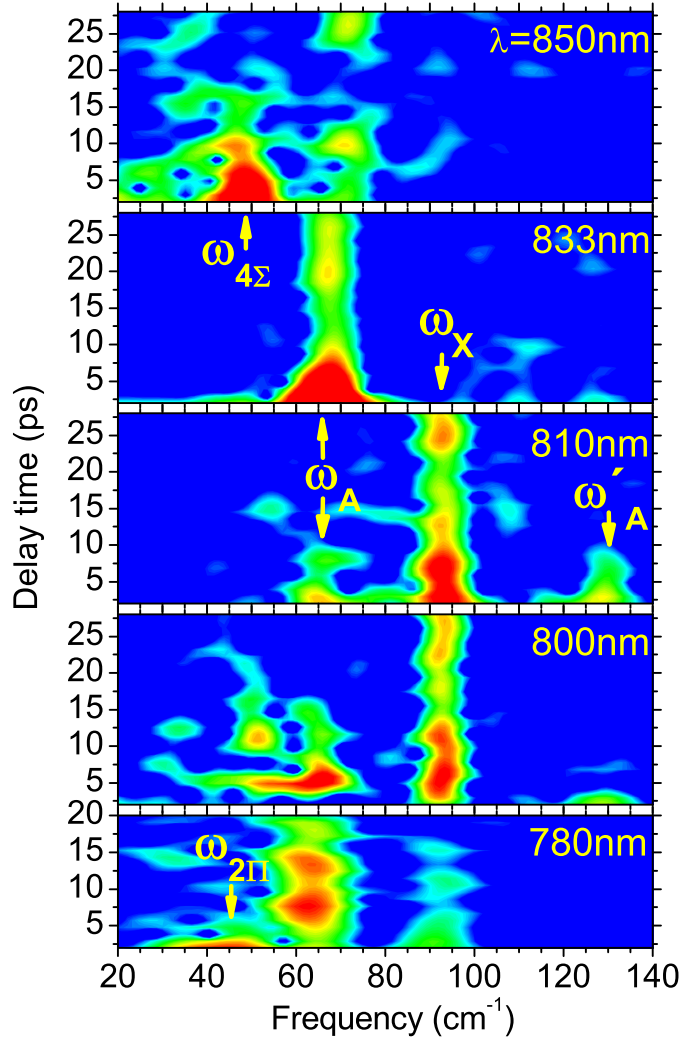}
\caption{(Color online) Overview over the spectrograms at characteristic excitation laser wavelengths.
\label{fig:Allspec}
}
\end{center}
\end{figure}

The time evolution of frequencies $\omega^{810nm}_X$, $\omega^{810nm}_A$ and $\omega^{'810nm}_A$ at $\lambda =
810\,$nm can be followed in the corresponding spectrogram in Fig.~\ref{fig:Allspec}. The ground state
component $\omega^{810nm}_X\approx 92\,$cm$^{-1}$ is visible during the entire scan and does not
reveal any significant frequency changes. The frequencies $\omega^{810nm}_A$ and $\omega^{'810nm}_A$,
however, are observable in the spectrogram only during the first 10\,ps and do not
display any revivals at later times. 
Again, this dynamics is associated with the K$_2$
desorbing from the droplets upon electronic excitation. It is interesting to note that,
in contrast to the amplitudes of components $\omega^{810nm}_A$ and $\omega^{'810nm}_A$,
no significant damping of the wave packet motion in the $X$-state excited by RISRS is observed.
Hence, no significant desorption of K$_2$ from the
helium droplets occurs as long as only vibrational states of the electronic ground state $X$
are redistributed. This observation agrees well with the picture of dopant molecules
desorbing from the surface only due to strong evaporation caused by a sudden perturbation
upon spatial expansion of the valence electronic distribution due to the laser induced transition~\cite{Vongehr:2003}. The
energy deposited into the droplet clearly shows up as intense blue-shifted sidebands when
probing alkali dimer transitions by means of fluorescence absorption spectroscopy.
By fitting an exponential
decay function to the amplitudes of the components $\omega^{810nm}_A$ and $\omega^{'810nm}_A$ in the same way as illustrated
in Fig.~\ref{fig:FTEvo} (b) at $\lambda =800\,$nm one obtains a time constant
$\tau^{810nm}_{\omega_A}=6.6(5)\,$ps for both $\omega^{810nm}_A$ and $\omega^{'810nm}_A$.
\cite{Stienkemeier2:1995,Higgins:1998}.

\paragraph{\textbf{800\,nm}}
The spectrogram at the excitation wavelength $\lambda = 800\,$nm (Fig.~\ref{fig:Allspec}) displays the time
evolution of the frequency components $\omega_X^{800nm}$, $\omega_A^{800nm}$ and $\omega^{'800nm}_{A}$, as
for $\lambda = 810\,$nm. The time evolution of the corresponding amplitudes is depicted in
Fig.~\ref{fig:FTEvo} (b). The ground state component $\omega_X^{800nm}$ remains largely unchanged during
the shown time interval. The slight increase in amplitude at $t_{delay}\approx 22\,$ps which is also
observed at $\lambda = 810\,$nm may reflect a revival of the wave packet motion. 
The amplitudes of components $\omega_A^{800nm}$ and $\omega^{'800nm}_{A}$, however, now behave differently
than at $\lambda = 810\,$nm. While the weaker component $\omega^{'800nm}_{A}\approx 130\,$cm$^{-1}$ disappears
with a time constant $\tau^{800nm}_{\omega'_A}=3.2(7)\,$ps, $\omega_{A}^{800nm}\approx 65\,$cm$^{-1}$
only reaches its maximum amplitude around 5\,ps. The following decay has a time constant $\tau^{800nm}_{\omega_A}=8(1)\,$ps.

This means that during the desorption process the FC window for the probe step into the ionic state
is shifted. In the beginning, transitions take place at the outer turning point as well as at the inner turning 
point of the $A$-state potential. As the influence of the helium diminishes upon desorption,
the FC-window shifts to the outer turning point. After desorption is complete, ionization
of the $A$-state is suppressed. Thus a transition from the excitation scheme of type I to the one
of type II is taking place at this wavelength for K$_2$ on helium droplets. Besides the shift of the
frequency $\omega_A$ at $\lambda =833\,$nm, the opening and closing of FC-windows observed at
$\lambda =800\,$nm is a clear indication for the dynamics induced by the desorption process of K$_2$ off
the helium droplets.

\paragraph{\textbf{780\,nm}}
Around 780\,nm, the recorded oscillation of the ion signal has a period changing from
$T_{A_1}^{780nm}=529\,$fs ($\omega_{A_1}^{780nm}=63\,$cm$^{-1}$) to the value observed in the
gas phase, $T_{A_2}^{780nm}=537\,$fs
($\omega_{A_1}^{780nm}=62.1\,$cm$^{-1}$), within the first 20\,ps. The phase is found to be
$\Phi_A^{780nm}= 0.1\,\pi$ (28\,fs). Thus, this wave packet motion occurs in the $A$-state according
to the type II scheme, which is also observed in gas-phase
experiments~\cite{Nicole:1999}. The changing oscillation period is again attributed to
the desorption dynamics. However, in contrast to the behavior at 833\,nm, in this case
$T_{A_1}^{780nm}<T_{A_2}^{780nm}$ as shown in Fig.~\ref{fig:spacing}. 
The amplitude of the component $\omega_A^{780nm}$ evolves differently at 780\,nm than at 
previously discussed wave lengths. From 0-8\,ps it rises by a factor two and decays slowly at later times.
The FC-window seems not to shift sufficiently during the desorption for significantly changing
the detection efficiency of the $A$-state wavepacket. Presumably, this is due to the fact that
the inner part of the K$_2$ dimer potential where the transition takes place at this wave length
is less sensitive to small perturbations by the external environment. 

Besides the dominating frequency component $\omega_A^{780nm}$ the spectrogram of Fig.~\ref{fig:Allspec}
displays the time evolution of $\omega_X^{780nm}\approx 45\,$cm$^{-1}$ and $\omega_{2\Pi}^{780nm}\approx 92\,$cm$^{-1}$.
The latter frequency reflects the wave packet motion excited in the $2\Pi$ state by the transition
sequence of type III. In the gas phase, this oscillation dephases within ca. 2\,ps due to the
anharmonicity of the potential. From fitting an exponential decay function to our data we find an exponential decay constant
$\tau^{780nm}_{\omega_{2\Pi}}=1.4(2)\,$ps, in agreement with the gas-phase observations.
The ground state component $\omega_X^{780nm}$ reveals a slow decrease which may however be
due to dephasing as it is observed at $\lambda =800\,$nm and $\lambda =810\,$nm. 

\paragraph{\textbf{850\,nm}}
In the range of 850\,nm the ionization transient displays a regular oscillation with a period
$T_{4\Sigma}^{850nm}\approx 700\,$fs ($\omega^{850nm}_{4\Sigma}\approx 48\,$cm$^{-1}$) and a phase
$\Phi_{4\Sigma}^{850nm}=1.9\,\pi$ (650\,fs). 
We interpret this oscillation as wave packet motion in the $4^1\Sigma_g^+$-potential detected by the type III
scheme involving state $4\Sigma$ instead of $2\Pi$. In quantum dynamics calculations of the ion yield at
the excitation wave length $\lambda = 840\,$nm, it was found that the state $4\Sigma$ is populated
but only weakly contributes to the ion yield originating predominantly from the type I process~\cite{Riedle:1995}.
Presumably the opposite is true as the wave length increases to $\lambda =850\,$nm. The average
vibrational splitting was found to be 52\,cm$^{-1}$ ($v'$=33-38) which roughly matches the
oscillation frequency $\omega^{850nm}_{4\Sigma}\approx 48\,$cm$^{-1}$ measured in our experiment.
As shown in Fig.~\ref{fig:Allspec}, the component $\omega^{850nm}_{4\Sigma}$ 
evolves an exponential decay with a time constant $\tau^{850nm}_{\omega_{4\Sigma}}=5.8(6)\,$ps
which agrees with the calculated dephasing time due to anharmonicity~\cite{Riedle:1995}. The 
amplitude of the weaker ground state component $\omega^{850nm}_X$ fluctuates around a constant value
indicating vanishing perturbation of the ground state dynamics.

\section{Summary}
The wavepacket motion observed by resonant and resonance-enhanced 3PI of K$_2$ on helium 
nanodroplets at various excitation wave lengths displays a number of characteristic
features as compared to analogous results of gas phase experiments.
Let us summarize the observations giving information about the influence of the helium
droplet on the wave packet dynamics. At $\lambda =833\,$nm, the strong oscillation
of the ion yield due to resonant 3PI is damped with a time constant $\tau =3.2(2)\,$ps
but a constant long-lived oscillation persists. On the same timescale the frequency of the
wave packet motion in the $A$-state shifts by 0.5\,cm$^{-1}$ from 65.8\,cm$^{-1}$ to the
gas phase value, 66.3\,cm$^{-1}$. 
At $\lambda =820\,$nm, no splitting of the maxima in the ion transient due to inward-outward
motion of the wave packet through the FC-window is measured contrary to gas phase observations.
At $\lambda =810\,$nm, enhanced RISRS is observed as opposed to negligible RISRS at
moderate laser intensities in the gas phase. The frequency shift $\Delta\omega_A$ at 
$\lambda =780\,$nm is comparable to the one at $\lambda =833\,$nm but has opposite sign.

The transient effect of the helium environment appears most prominently in the wave packet dynamics
observed according to the type I scheme. Here, the ionization efficiency depends sensitively on
the outer part of the potential curve which is particularly affected by the interaction with
the helium. This explains the strong amplitude damping and the shifting frequency at $\lambda =833\,$nm.
The ionization processes of type II to IV proceed via the inner part of the excited potential curves.
Therefore the influence of the surrounding helium is expected to be less visible.
In particular, the time varying perturbation by the helium as the K$_2$ dimers desorb from the droplet
surface causes different dynamics. 
As the ionization scheme gradually evolves from type I to type II for decreasing
wave lengths $\lambda = 833\rightarrow 780\,$nm, the competition of the two 
schemes is observable by the appearance and disappearance of the frequency
components connected with the different schemes, as observed at $\lambda = 800\,$nm. 
Consequently, the amplitude of $\omega_A$ at $\lambda = 780\,$nm (type II process) displays
only a weak variation with time. 

In contrast to these findings, the wave packets excited by RISRS
in the ground state experience no significant influence by the helium environment. The
vibrational frequency remains unshifted with respect to the gas phase, and the
amplitude of the detected wave packet motion in the ground state only weakly varies with time,
presumably reflecting a revival structure. Since no change of the valence electronic
distribution occurs upon RISRS, only coupling of the vibrational excitation to the helium bath might
induce perturbations of the wave packet motion. Since the latter is known to be much weaker
than the interaction with the helium droplet upon electronic excitation, no significant
perturbations are expected. 

In order to draw more quantitative conclusions from our observations quantum dynamics simulations
should be performed, introducing perturbed K$_2$ potential curves to mimic the influence of the helium
droplet. Also, the process of desorption of alkali atoms and molecules from helium droplets and, in
particular, the impact on the nuclear wave packet dynamics, is largely unexplored by theory. 

The experiments presented in this work clearly demonstrate that the influence of the helium matrix on the wave packet
dynamics of K$_2$ can be studied in real-time using fs pump-probe spectroscopy.
Surprisingly, this weak coupling quite dramatically
alters the photoionization process which demonstrates the sensitivity of Franck-Condon factors on
helium induced shifts of the potential energies. The time variation of the Franck-Condon windows
and of oscillation frequencies converging to their gas phase values
clearly reflects the dynamics of the process of desorption of the
K$_2$ dimers off the helium droplets, which proceeds on the timescale $3-8\,$ps. 

\section*{Acknowledgments}
Technical support with the laser system by V. Petrov
and the help with producing spectrograms by S. Rutz is gratefully acknowledged.
The work is financially supported by the DFG.
\section*{References}

\bibliographystyle{unsrt}
\bibliography{OlliBib}

\begin{thebibliography}{10}

\bibitem{Zewail:2000}
A.~H. Zewail.
\newblock {\em Journal of Physical Chemistry A}, 104:5660, 2000.

\bibitem{Zewail:1994}
A.~Zewail.
\newblock {\em Femtochemistry}.
\newblock World Scientific, Singapore, 1994, 1994.

\bibitem{Manz:1995}
J.~Manz and L.~Wöste, editors.
\newblock {\em Femtosecond Chemistry}.
\newblock VCH, Weinheim, 1995.

\bibitem{Chergui:1995}
M.~Chergui, editor.
\newblock {\em Femtochemistry}.
\newblock World Scientific, Singapore, 1995.

\bibitem{Apkarian:1999}
V.~A. Apkarian and N.~Schwentner.
\newblock {\em Chem. Rev.}, 99:1481, 1999.

\bibitem{Wan:1997}
C.~Wan, M.~Gupta, J.~Baskin, Z.~Kim, and A.~H. Zewail.
\newblock {\em Journal of Chemical Physics}, 106:4353, 1997.

\bibitem{Liu:1993}
Q.~Liu, J.-K. Wang, and A.~Zewail.
\newblock {\em Nature}, 364:427, 1993.

\bibitem{Lienau:1994}
C.~Lienau and A.~H. Zewail.
\newblock {\em Chemical Physics Letters}, 222:224, 1994.

\bibitem{Bargheer:1999}
M.~Bargheer, P.~Dietrich, K.~Donovang, and N.~Schwentner.
\newblock {\em Journal of Chemical Physics}, 111:8556, 1999.

\bibitem{Fushitani:2006}
M.~Fushitani, N.~Schwentner, M.~Schröder, and O.~Kühn.
\newblock {\em Journal of Chemical Physics}, 124:024505, 2006.

\bibitem{Guehr:2004}
M.~Gühr, H.~Ibrahim, and N.~Schwentner.
\newblock {\em Physical Chemistry Chemical Physics}, 6:5353, 2004.

\bibitem{Bargheer:2002}
M.~Bargheer, M.Y. Niv, R.B. Gerber, and N.~Schwentner.
\newblock {\em Physical Review Letters}, 89:108301, 2002.

\bibitem{Gonzalez:2002}
C.~R. Gonzalez, S.~Fernandez-Alberti, and J.~Echave.
\newblock {\em Journal of Chemical Physics}, 116:3343, 2002.

\bibitem{Benderskii:1999}
A.~V. Benderskii, R.~Zadoyan, N.~Schwentner, and V.~A. Apkarian.
\newblock {\em Journal of Chemical Physics}, 110:1542, 1999.

\bibitem{Comstock:2003}
M.~Comstock, V.~V. Lozovoy, and Marcos Dantus.
\newblock {\em Journal of Chemical Physics}, 119:6546, 2003.

\bibitem{Papanikolas:1995}
J.~M. Papanikolas, R.~M. Williams, P.~Kleiber, J.~L. Hart, C.~Brink, S.~D.
  Price, and S.~R. Leone.
\newblock {\em Journal of Chemical Physics}, 103:7269, 1995.

\bibitem{Baumert:1991}
T.~Baumert, M.~Grosser, R.~Thalweiser, and G.~Gerber.
\newblock {\em Physical Review Letters}, 67:3753, 1991.

\bibitem{Baumert:1992}
T.~Baumert, V.~Engel, C.~Meier, and G.~Gerber.
\newblock {\em Chemical Physics Letters}, 191:639, 1992.

\bibitem{Riedle:1995}
R.~de~Vivie-Riedle, B.~Reischl~S. Rutz, and E.~Schreiber.
\newblock {\em Journal of Physical Chemistry A}, 99:16829, 1995.

\bibitem{Riedle:1996}
R.~de~Vivie-Riedle, K.~Kobe, J.~Manz, W.~Meyer, B.~Reischl, S.~Rutz,
  E.~Schreiber, and L.~Wöste.
\newblock {\em Journal of Physical Chemistry A}, 100:7789, 1996.

\bibitem{Rodriguez:1993}
G.~Rodriguez and J.~G. Eden.
\newblock {\em Chemical Physics Letters}, 205:371, 1993.

\bibitem{Blanchet:1995}
V.~Blanchet, M.~A. Bouchene, O.~Cabrol, and B.~Girard.
\newblock {\em Chemical Physics Letters}, 233:491, 1995.

\bibitem{Rutz:1996}
Soeren Rutz, Regina de~Vivie-Riedle, and Elmar Schreiber.
\newblock {\em Physical Review A}, 54:306, 1996.

\bibitem{Rutz:1997}
Soeren Rutz and Elmar Schreiber.
\newblock {\em Chemical Physics Letters}, 269:9, 1997.

\bibitem{Schwoerer:1997}
H.~Schwoerer, R.~Pausch, M.~Heid, V.~Engel, and W.~Kiefer.
\newblock {\em Journal of Chemical Physics}, 107:9749, 1997.

\bibitem{Pausch:1999}
R.~Pausch, M.~Heid, T.~Chen, W.~Kiefer, and H.~Schwoerer.
\newblock {\em Journal of Chemical Physics}, 110:9560, 1999.

\bibitem{Nicole:1999}
C.~Nicole, M.~A. Bouch\`ene, C.~Meier, S.~Magnier, E.~Schreiber, and B.~Girard.
\newblock {\em Journal of Chemical Physics}, 111:7857, 1999.

\bibitem{Toennies:2004}
J.~P. Toennies and A.~F. Vilisov.
\newblock {\em Angewandte Chemie}, 43:2622--2648, 2004.

\bibitem{Stienkemeier:2006}
Frank Stienkemeier and Kevin Lehmann.
\newblock {\em J.~Phys.~B}, 39:R127, 2006.

\bibitem{Dalfovo:1994}
F.~Dalfovo.
\newblock {\em Zeitschrift für Physik D}, 29:61--66, 1994.

\bibitem{Ancilotto:1995}
F.~Ancilotto, G.~DeToffol, and F.~Toigo.
\newblock {\em Physical Review B}, 52:16125--16129, 1995.

\bibitem{Stienkemeier2:1995}
F.~Stienkemeier, J.~Higgins, W.~E. Ernst, and G.~Scoles.
\newblock {\em Physical Review Letters}, 74:3592--3595, 1995.

\bibitem{Stienkemeier:1995}
F.~Stienkemeier, W.~E. Enrst, J.~Higgins, and G.~Scoles.
\newblock {\em Journal of Chemical Physics}, 102:615--617, 1995.

\bibitem{Higgins:1998}
J.~Higgins, C.~Callegari, J.~Reho, F.~Stienkemeier, W.~E. Ernst, M.~Gutowski,
  and G.~Scoles.
\newblock {\em Journal of Physical Chemistry A}, 102:4952--4965, 1998.

\bibitem{HigginsThesis:1998}
J.~P. Higgins.
\newblock {H}elium {C}luster {I}solation {S}pectroscopy.
\newblock PhD thesis, Princeton University, 1998.

\bibitem{Higgins:2000}
J.~P. Higgins, J.~Reho, F.~Stienkemeier, W.~E. Ernst, K.~K. Lehmann, and
  G.~Scoles.
\newblock Spectroscopy in, on, and off a beam of superfluid helium
  nanodroplets.
\newblock In R.~Campargue, editor, {\em Atomic and Molecular Beams: the state
  of the Art 2000}, pages 723--754. Springer, 2000.

\bibitem{Stienkemeier:1996}
F.~Stienkemeier, J.~Higgins, C.~Callegari, S.~I. Kanorsky, W.~E. Ernst, and
  G.~Scoles.
\newblock {\em Zeitschrift für Physik D}, 38:253--263, 1996.

\bibitem{Callegari:1998}
C.~Callegari, J.~Higgins, F.~Stienkemeier, and G.~Scoles.
\newblock {\em Journal of Physical Chemistry A}, 102:95--101, 1998.

\bibitem{Mudrich:2004}
M.~Mudrich, O.~Bünermann, F.~Stienkemeier, O.~Dulieu, and M.~Weidemüller.
\newblock {\em European Physical Journal D}, 31:291--299, 2004.

\bibitem{Reho2:2001}
J.~Reho, J.~Higgins, and K.~K. Lehmann.
\newblock {\em Faraday Discussion}, 118:33--42, 2001.

\bibitem{Takayanagi:2003}
T.~Takayanagi and M.~Shiga.
\newblock {\em Chemical Physics Letters}, 372:90, 2003.

\bibitem{Braun:2004}
A.~Braun and M.~Drabbels.
\newblock {\em Physical Review Letters}, 93:253401, 2004.

\bibitem{Loginov:2005}
E.~Loginov, D.~Rossi, and M.~Drabbels.
\newblock {\em Physical Review Letters}, 95:163401, 2005.

\bibitem{Grebenev:1998}
S.~Grebenev, J.~P. Toennies, and A.~F. Vilesov.
\newblock {\em Science}, 279:2083--2085, 1998.

\bibitem{HartmannPRL:1996}
M.~Hartmann, F.~Mielke, J.~P. Toennies, and A.~F. Vilesov.
\newblock {\em Physical Review Letters}, 76:4560, 1996.

\bibitem{Higgins:1996}
J.~Higgins, C.~Callegari, J.~Reho, F.~Stienkemeier, W.~E. Ernst, K.~K. Lehmann,
  M.~Gutowski, and G.~Scoles.
\newblock {\em Science}, 273:629--631, 1996.

\bibitem{Schulz:2004}
C.~P. Schulz, P.~Claas, D.~Schumacher, and F.~Stienkemeier.
\newblock {\em Physical Review Letters}, 92:013401, 2004.

\bibitem{Magnier:1996}
S.~Magnier and P.~Mill\'{e}.
\newblock {\em Physical Review A}, 54:204, 1996.

\bibitem{Lyyra:1992}
A.~M. Lyyra, T.~T. Luh, L.~Li, H.~Wang, and W.~C. Stwalley.
\newblock {\em Journal of Chemical Physics}, 97:2, 1992.

\bibitem{Guehr:2002}
M.~Gühr, M.~Bargheer, P.~Dietrich, and N.~Schwentner.
\newblock {\em Journal of Physical Chemistry A}, 106:12002, 2002.

\bibitem{Vongehr:2003}
S.~Vongehr and V.~V. Kresin.
\newblock {\em Journal of Chemical Physics}, 119:11124--11129, 2003.

\end{thebibliography}
\end{document}